# Strange correlations between remote nodes in networks comprising chaotic links

Julia Manasson[1] and Vladimir A. Manasson[2]

[1] New York University School of Medicine, New York, NY

[2] Sierra Nevada Corporation, Irvine, CA

*Abstract*

We studied correlations between different nodes in small electronic networks with active links operating as jitter generators. Unexpectedly, we found that under certain conditions signals from the most remote nodes in the networks correlate stronger than signals from all of the other coupled nodes. The phenomenon resembles selective remote correlation between electrons in Cooper's pairs or entangled particles.

*1. Introduction*

Chaotic systems and networks sometimes exhibit phenomena, which at first glance may appear strange and counterintuitive, like synchronization of chaotic systems [1], or suppression of chaos by chaos [2]. The term "deterministic chaos" [3] in itself sounds strange. In this communication, we report another counterintuitive behavior of circuits comprising active links, which we named strange remote correlation (SRC).

The nodes in the networks can be connected in different ways (via different links). For any two nodes $x, y$, there always exist the shortest connections ($SC_{x,y}$), which correspond to the minimal number of links between these nodes. The longer $SC_{x,y}$, the more remote the nodes $x, y$. The most remote nodes, $(x, y)_r$, are those for which $SC_r$ is longer than $SC_{x,y}$ for all other nodes $x, y$ in a given network.

We studied correlation among different nodes in small circuit networks, where all links are active and operate as jitter generators. We found that under certain conditions the correlation between signals from the two most remote nodes, $SC_r$, is significantly stronger than the correlation between signals from all other couples of nodes with shorter $SCs$.

In this communication we provide a detailed description of two networks where we observed SRC. The network designs and architectures were proposed in references [4, 5]. One network comprises 32 active links and the other 16 active links. The active links operating as jitter generators resemble the circuit described in reference [6] with the modification that the negative feedback photodiodes in the original circuit are replaced with regular diodes.

*2. Basic circuit elements*

The simplest circuit comprising all three basic elements we used to assemble the networks is shown in Fig.1.

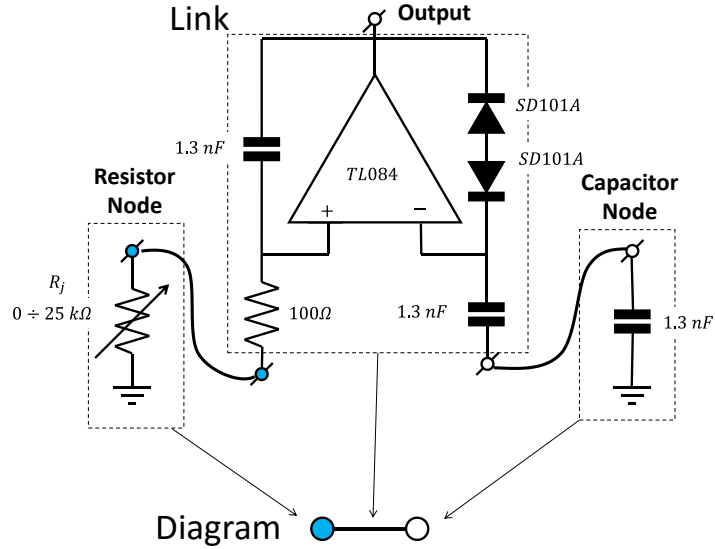

Fig. 1. Simplest circuit comprising all basic elements: resistor node (blue terminals), capacitor node (white terminals), and active jitter-generator link.

It comprises only two nodes—one is grounded via a variable resistor (represented by blue terminals in the circuit and a blue circle in the link diagram) and the other is grounded via a capacitor (represented by white terminals in the circuit and a white circle in the link diagram). In the following discussion we will refer to them as the "resistor node" and the "capacitor node". The nodes are coupled by an active link, which consists of a differential amplifier with capacitive positive feedback and nonlinear negative feedback. The latter is represented by two diodes connected in series in mutually opposite directions. We will refer to this link as the "jitter-generator link".

We recorded time waveforms and phase portraits using a two-channel digital oscilloscope. $V_R$ and $V_C$ were respectively measured at the resistor node and capacitor node terminals. Samplings were taken with equal time intervals $\Delta t$. Each waveform length consisted of $N = 4096$ sampling points. We calculated auto-correlation functions $AC$ between different pulse strings of the same waveform, and cross-correlation functions $CC$ between pulse strings of waveforms taken at different resistor nodes. Each correlation function was calculated at $N/2$ points with the corresponding time delays $p\Delta t$, $p = 0, 1 \ldots \frac{N}{2} - 1$. The lengths of the waveform strings used for calculations consisted of $N/2$ points.

The correlation functions were calculated using formulas from [7]. The auto-correlation functions at time delay $p\Delta t$ were calculated as

$$AC_p = \left(\sum_{i=0}^{\frac{N}{2}} v_i v_{i+p}\right) \bigg/ \left(\sum_{i=0}^{\frac{N}{2}} v_i^2\right) \quad (1),$$

where $v_i$ and $v_{i+p}$ are normalized voltages measured at time instants $t_i$ and $t_{i+p}$,

$$v_i = (V_i - \langle V \rangle)/max(V_i - \langle V \rangle),$$

and brackets ⟨ ⟩ signify the average values. The cross-correlation functions between $j$-th and $k$-th resistor nodes at time delay $p\Delta t$ were calculated as

$$CC_{p,k,j} = \left(\sum_{i=0}^{\frac{N}{2}} v_{i,k} v_{i+p,j}\right) \Big/ \left(\sqrt{\sum_{i=0}^{\frac{N}{2}} v_{i,k}^2 \sum_{i=0}^{\frac{N}{2}} v_{i,j}^2}\right) \quad (2).$$

A typical waveform produced by the circuit in Fig. 1 and measured at the operational amplifier output is shown in Fig. 2. The pattern represents series of irregular pulses (jitter).

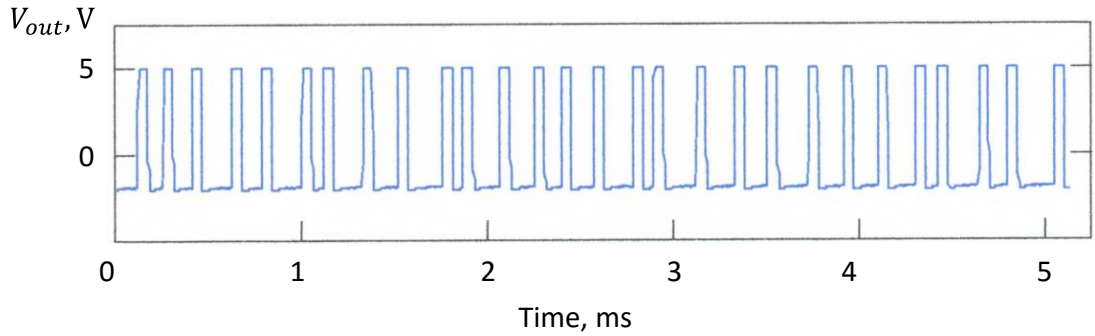

Fig. 2. Typical jitter waveform produced by circuit shown in Fig. 1.

The circuit produces a continuous power spectrum (Fig. 3) with the highest peak corresponding to the principal frequency, which can barely be differentiated from the background noise.

We also recorded phase portraits — parametric curves depicting how two potentials $(V_x(t), V_y(t))$ measured at different circuit nodes $(x, y)$ relate in time. A typical phase portrait $(V_R(t), V_C(t))$ recorded at the resistor terminal and capacitor terminal for the circuit in Fig. 1 is shown in Fig. 4. It is also noisy, in agreement with the jitter-waveform and the power spectrum.

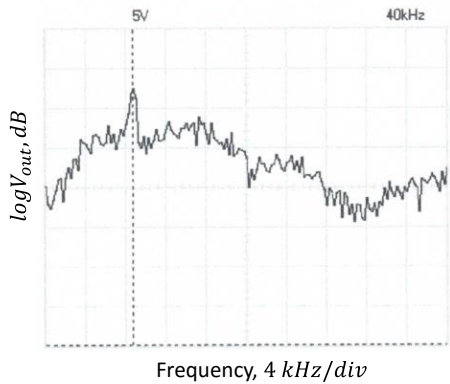

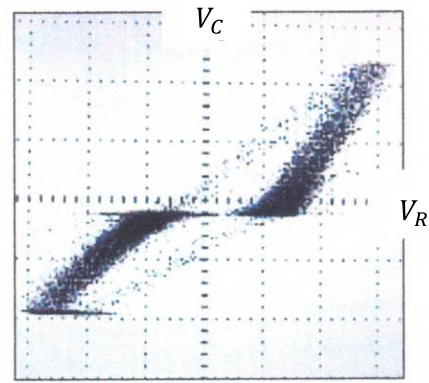

Fig. 3. Typical power spectrum measured at the amplifier output of the circuit in Fig.1.

Fig. 4. Typical phase portrait, $(V_C(t), V_R(t))$, for the circuit in Fig.1.

A typical auto-correlation function calculated for the circuit in Fig. 1 is shown in Fig. 5. The autocorrelation function, $AC$, quickly drops from 100% to values that never exceed 45%.

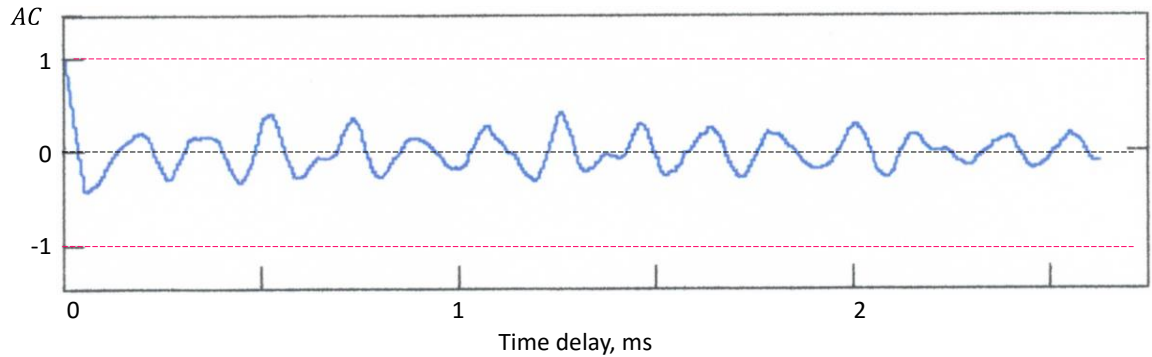

Fig. 5. Auto-correlation function for the circuit in Fig. 1.

## 3. Hypercube

The first network we will consider in this paper has the topology of a four-dimensional cube, and we will refer to it as "hypercube" (Fig. 6). Eight hypercube nodes are grounded via variable resistors $R_j$ (represented by colored circles), and eight nodes are grounded via capacitors (represented by white circles). The nodes are connected by thirty-two jitter-generator links. To minimize distortions caused by the oscilloscope, most of measurements were made at the resistors nodes. For convenience, the latter are marked by different colors and different numbers ($j = 1, 2, ..., 8$).

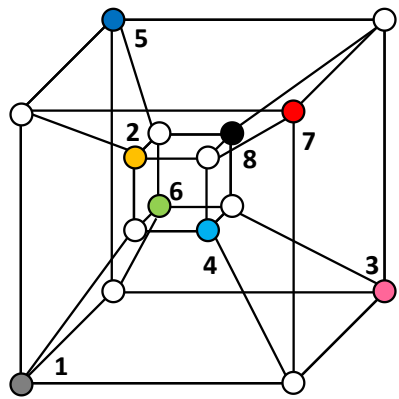

Fig. 6. Hypercube network diagram. Capacitor nodes are represented by white circles. Resistor nodes are represented by colored circles. Jitter-generator links are represented by "hypercube" edges.

### 3.1. Regular oscillations in hypercube

In the hypercube network, the jitter is suppressed for some sets of $R_j$ values. One example of stable oscillations obtained with the resistor set shown in Fig. 7 (left) is described in this section.

An example of typical phase portraits, $(V_1(t), V_j(t)), j = 2, ..., 8$, measured at resistor node 1 and all other resistor nodes is shown in Fig. 7. We observed four different stable phase portraits. During measurements, the potentials $V_3(t) \approx V_6(t)$ and $V_7(t) \approx V_8(t)$ 6, and the corresponding phase portraits were almost identical.

The corresponding cross-correlation functions are shown in Fig. 8. Their maximal values periodically approach almost 100%, $(CC_{1,j})_{max} \sim 1$ for all curves $j = 2, ..., 8$.

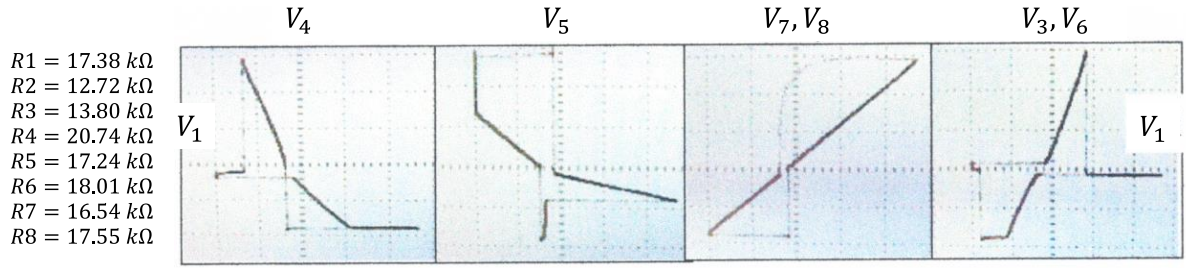

Fig. 7. Resistor values (left) and phase portraits, $(V_1(t), V_j(t))$, recorded at node 1 and all other nodes $j = 2, .., 8$.

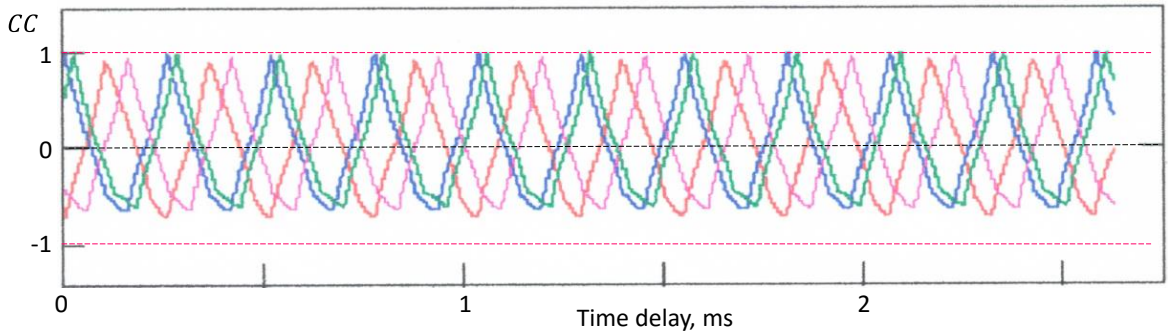

Fig. 8. Hypercube cross-correlation functions, $CC_{1,j}, j = 2, ..., 8$, for the resistor set in Fig. 7. Curve colors correspond to resistor-node colors in Fig. 6.

### 3.2. Strange remote correlations in hypercube

For certain $R_j$ values, the hypercube network demonstrated chaotic behavior with various correlation patterns. Here we are interested in SRC. One case of SRC is described below.

The data was recorded for $R_j$ values shown in Fig. 9 (left). A typical phase portrait, $(V_1(t), V_8(t))$, recorded at node 1 and the most remote node 8 is shown in Fig. 9a. It is noisy. However, the phase portraits recorded at node 1 and all other nodes, $(V_1(t), V_j(t)), j = 2, .., 7$, are much noisier. One of them is shown in Fig. 9b.

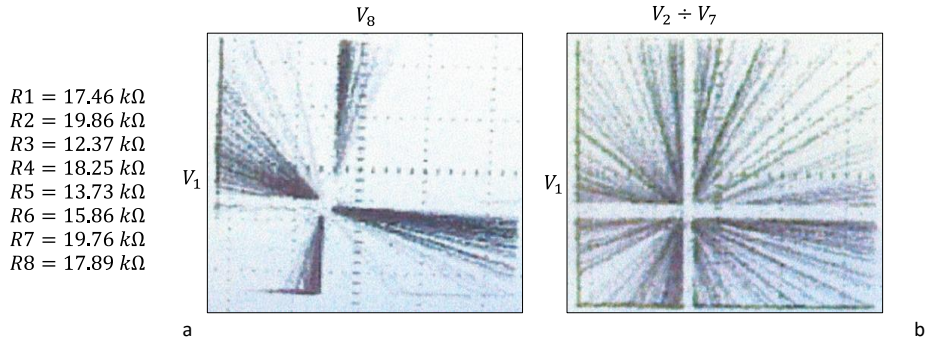

Fig. 9. Typical phase portraits: (a) $(V_1(t), V_8(t))$ measured at resistor nodes 1 and 8; (b) $(V_1(t), V_j(t)), j \neq 8$ measured at node 1 and nodes 2 through 7.

The cross-correlation functions are shown in Fig. 10. They are in concert with the phase portraits shown in Fig. 9.

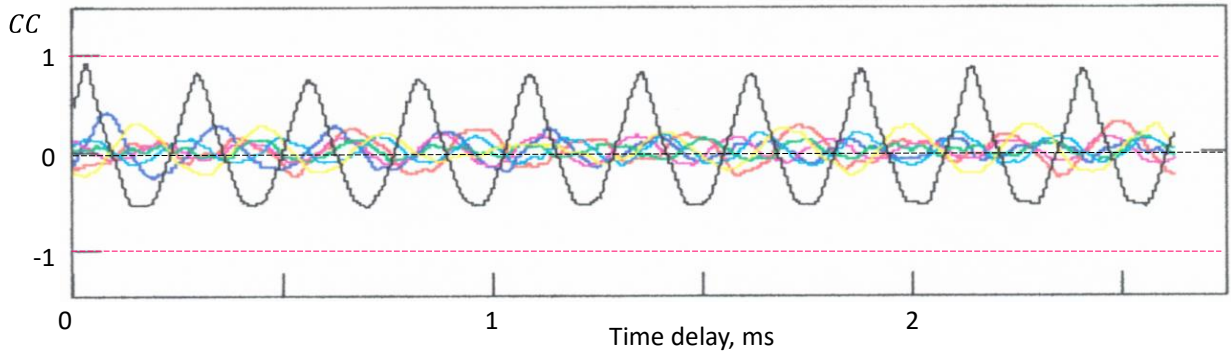

Fig. 10. Cross-correlation functions for the resistor set shown in Fig. 9. Colors correspond to node colors shown in Fig. 6.

The cross-correlations between node 1 and nodes 2 through 7 never exceed 40% $(CC_{1,j})_{max} < 0.4, j = 2 \ldots 7$ (colored curves), while the cross-correlation between node 1 and its *most remote* node (node 8) reaches values that exceed 90%, $(CC_{1,8})_{max} \sim 0.9$ (black curve).

### 4. *SRC in a smaller circuit*

In this section we describe SRC in a smaller network, which is shown in Fig. 11. It comprises only four resistor nodes (colored circles) and eight capacitor nodes (white circles), which are connected via sixteen jitter-generator links. One of the resistor sets, which results in SRC and its typical phase portraits are shown in the same figure. The phase-portrait frame colors correspond to the resistor node colors.

The phase portrait, $(V_1(t), V_2(t))$, measured at resistor nodes 1 and 2 and the phase portrait, $(V_1(t), V_3(t))$, measured at nodes 1 and 3 demonstrate a significant amount of noise. Conversely, the phase portrait, $(V_1(t), V_4(t))$, measured at node 1 and its *most remote* resistor node (node 4), depicts regular oscillations.

The cross-correlation functions $CC_{1,j}$, $j = 2 \ldots 4$ are shown in Fig. 12. The curve colors correspond to the node colors shown in Fig. 11.

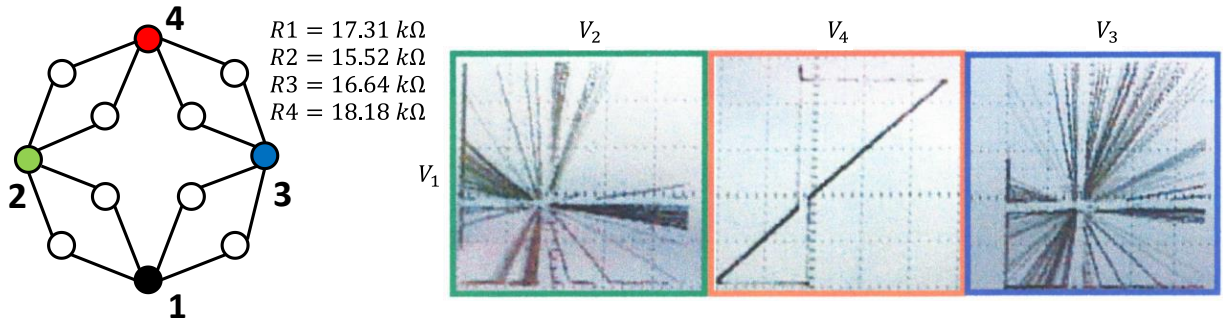

Fig. 11. Circuit diagram (left), resistor values, and phase portraits $\left(V_1(t), V_j(t)\right), j = 2, 4, 3$ (from left to right) measured at corresponding resistor nodes.

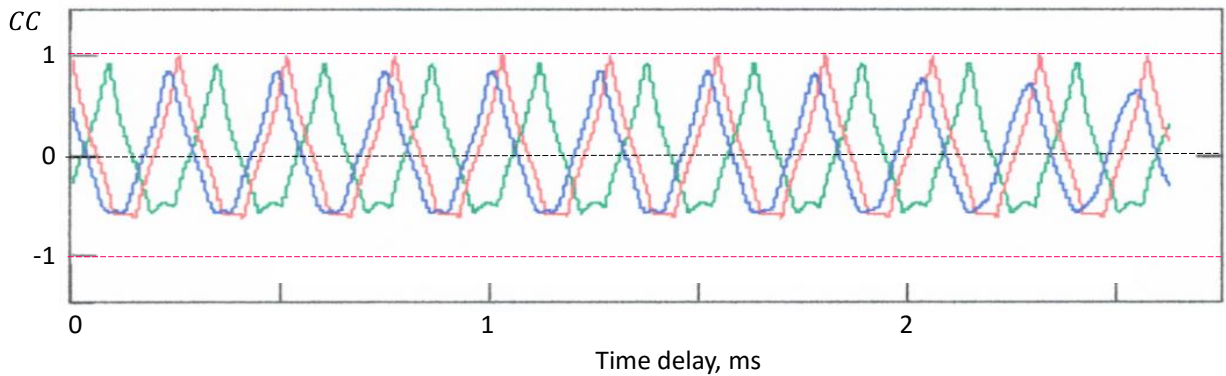

Fig. 12. Cross-correlation functions $CC_{1,j}$, $j = 2 \ldots 4$, for the circuit shown in Fig. 11. Colors correspond to node colors.

The maximal correlation between resistor nodes 1 and 2 is $(CC_{1,2})_{max} \sim 0.95$ (green curve) and between nodes 1 and 3 is $(CC_{1,3})_{max} \sim 0.87$ (blue curve), while the correlation between the most remote nodes 1 and 4 is $(CC_{1,4})_{max} \sim 1$ (red curve), reaching 100%.

## 5. Brief discussion

Intuitively, we would expect that correlation strength would monotonically decrease as a function of the distance $SC$ between nodes. In linear networks, where all connections are of $SC$ type, this is indeed the case. However, in the networks described above, the nodes are connected in series as well in parallel. Most likely the latter is responsible for the SRC phenomenon.

The observed SRC in some ways resembles *selective* quantum correlation between particles at a distance. For example, spin-correlation in superconductors may occur between two remotely separated electrons (the Cooper pairs [8]) even though there are a number of non-correlated electrons located at closer distances. The same is true of quantum entanglement [9], where the entangled particles are not necessarily located in proximity to each other and do not entangle with identical particles from their immediate neighborhood.